\documentclass[prb,twocolumn,showpacs,preprintnumbers,amsmath,amssymb]{revtex4}

\usepackage{graphicx}
\usepackage{dcolumn}
\usepackage{bm}

\begin{document}


\title{Electronic properties of single-crystalline Fe$_{1.05}$Te and Fe$_{1.03}$Se$_{0.30}$Te$_{0.70}$}

\author{G. F. Chen}
\author{Z. G. Chen}
\author{J. Dong}
\author{W. Z. Hu}
\author{G. Li}
\author{X. D. Zhang}
\author{P. Zheng}
\author{J. L. Luo}
\author{N. L. Wang}

\affiliation{Beijing National Laboratory for Condensed Matter
Physics, Institute of Physics, Chinese Academy of Sciences,
Beijing 100190, People¡¯s Republic of China}


\begin{abstract}
We report on a comprehensive study of the transport, specific
heat, magnetic susceptibility and optical spectroscopy on single
crystal of Fe$_{1.05}$Te. We confirm that Fe$_{1.05}$Te undergoes
a first-order phase transition near 65 K. However, its physical
properties are considerably different from other parent compounds
of FeAs-based systems, presumably attributed to the presence of
excess Fe ions. The charge transport is rather incoherent above
the transition, and no clear signature of the gap is observed
below the transition temperature. Strong impurity scattering
effect exists also in Se-doped superconducting sample
Fe$_{1.03}$Se$_{0.30}$Te$_{0.70}$, leading to a relatively low
T$_c$ but a rather high upper critical field.

\end{abstract}



\maketitle

The recent discovery of superconductivity in Fe- and Ni-based
transition metal oxypnictides has triggered tremendous interest in
searching for new Fe-based superconductors with similar PbO-type
tetrahedral layers. \cite{Kamihara08} $\alpha$-FeSe(Te) belongs to
this tetragonal family with the simplest crystal structure. It
comprises only a continuous stacking of tetrahedral FeSe(Te)
layers along the c-axis. Superconductivity with transition
temperature T$_c$ up to 15 K was obtained on Fe$_{1+x}$(Se,Te)
system at ambient pressure.\cite{Wu01,Mao,Wu02} The T$_c$ can go
up to 27 K at a pressure of 1.48 GPa.\cite{Mizuguchi} First
principle calculations\cite{Sybedi} on stoichiometric FeSe
indicate that the electron-phonon coupling cannot explain the
superconductivity at such a high transition temperature, and FeSe
is in the category of unconventional superconductivity. The
calculated Fermi surface (FS) structures of FeS, FeSe and FeTe are
very similar to that of FeAs based superconductors, with
cylindrical hole sections at the zone center and electron sections
at the zone corner. Those Fermi surfaces are separated by a
two-dimensional nesting wave vector at ($\pi,\pi$).
Spin-density-wave (SDW) ground state is obtained due to the
substantial FS nesting effect. In particular, in going from FeSe
to FeTe, the strength of the SDW instability is strongly enhanced.

Experimentally, it is difficult to synthesize the stoichiometric
$\alpha$-FeSe(Te), and excess Fe is always present in synthesized
compounds.\cite{Wu01,Mao,Wu02,Fruchart,Bao,Li} Those excess Fe
ions occupy randomly on the Fe(2) sites as in Fe$_2$As compound or
the Li sites in LiFeAs compound (both have Cu$_2$Sb type
structure) at a low partial filling\cite{Bao,Li,Pitcher,Zhang}.
The superconductivity was found to appear in wide range of
compositions Fe$_{1+y}$Se$_x$Te$_{1-x}$(0$<$x$\leq$1),\cite{Mao,
Wu02} but the Se-free Fe$_{1+y}$Te undergoes a structure
distortion along with the establishment of a long range SDW order
near 65K.\cite{Fruchart,Bao,Li} This is consistent with the
expectation of the first principle density functional calculation
showing higher strength of SDW instability for FeTe.\cite{Sybedi}
However, a complex incommensurate long range antiferromagnetic
(AF) order is formed.\cite{Bao} With the reduction of the excess
Fe, the AF order tends to become commensurate, but the moments
rotate 45 degrees relative to the moment direction found in other
FeAs-based AF spin structures (along the orthorhombic long (a)
axis).\cite{Bao,Li} More recent density functional
calculations\cite{Zhang} indicate that the excess Fe is in a
valence state near Fe$^{+}$ and therefore donates electron to the
system. Furthermore, the excess Fe has a large magnetic moment and
interacts with the magnetism of FeTe layers, resulting in a
complex magnetic order.

Due to the interaction of the magnetic moment of excess Fe with
the itinerant electrons of FeTe layer, we can expect that the
electronic properties would be substantially affected as well. It
is of great interest to investigate the electronic properties of
this material and compare them with LaFeAsO and AFe$_2$As$_2$
(A=Ba,Sr) systems. Here we report on a comprehensive study of the
transport, specific heat, magnetic susceptibility and optical
spectroscopy on single crystals of Fe$_{1.05}$Te and
Fe$_{1.03}$Se$_{0.30}$Te$_{0.70}$. We confirm that Fe$_{1.05}$Te
undergoes a first-order phase transition near 65 K. Above
T$_{SDW}$, dc resistivity shows a semiconducting like behavior. In
accord with this non-metallic property, the optical conductivity
is rather flat with an absence of Drude component. This suggests
that there is almost no well-defined quasiparticle with
sufficiently long life time. The charge transport is rather
incoherent. Below T$_{SDW}$, a small Drude weight develops from
the incoherent background. But unlike LaFeAsO and SrFe$_2$As$_2$,
there is no clear sign of gap opening from optical, Hall
coefficient and low-T specific heat measurement results. This is
compatible with the disappearance of commensurate SDW caused by
the size mismatch between the electron and hole Fermi surfaces
induced by excess Fe as suggested by recent density functional
calculations.\cite{Zhang} The present work provides strong
evidence that the excess Fe moments strongly scatter the charge
carriers in FeAs layers. The superconducting properties of a
Fe$_{1.03}$Se$_{0.30}$Te$_{0.70}$ single crystal are also
investigated. A surprisingly high upper critical field is
observed, again yielding evidence for strong impurity scattering
effect in the superconducting sample.

Large single crystals of Fe$_{1.05}$Te and
Fe$_{1.03}$Se$_{0.30}$Te$_{0.70}$ have been grown by the Bridgeman
technique. The starting compositions were selected as FeTe and
FeSe$_{0.30}$Te$_{0.70}$, respectively. The mixtures of Fe, Te(Se)
were grounded thoroughly and sealed in an evacuated quartz tube.
The tube was heated to 920 $^{\circ}C$ and cooled slowly to grow
single crystals. The obtained crystals were checked by X-ray
diffraction (XRD) and their compositions were determined by
Inductively Coupled Plasma (ICP) analysis. The resistivity was
measured by the standard 4-probe method. The DC magnetic
susceptibility was measured with a magnetic field of 1 T. The Hall
coefficient measurement was done using a five-probe technique. The
specific heat measurement was carried out using a thermal
relaxation calorimeter. Those measurements were performed down to
1.8 K in a Physical Property Measurement System(PPMS) of Quantum
Design Company. The optical reflectance measurements were
performed on Bruker IFS 66v/s spectrometer on newly cleaved
surfaces (ab-plane) of those crystals up to 25000 cm$^{-1}$. An
\textit{in situ} gold and aluminium overcoating technique was used
to get the reflectivity R($\omega$). The real part of conductivity
$\sigma_1(\omega)$ is obtained by the Kramers-Kronig
transformation of R($\omega$).

\begin{figure}
\centerline{\includegraphics[width=8.5cm]{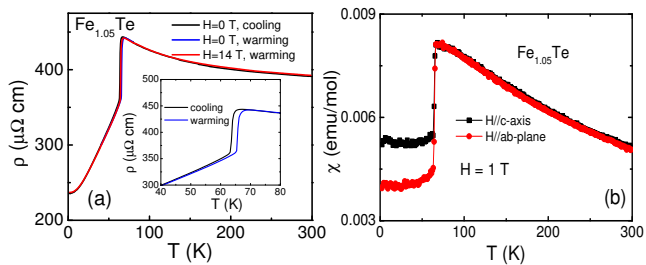}}%
\caption{(Color online) (a) The in-plane resistivity $\rho_{ab}$
for Fe$_{1.05}$Te in zero field and 14T with H$\parallel$c-axis,
respectively. (b) Magnetic susceptibility of Fe$_{1.05}$Te as a
function of temperature with H$\parallel$ab plane and
H$\parallel$c-axis, respectively.}
\end{figure}

Figure 1(a) shows the temperature dependence of in-plane
resistivity $\rho_{ab}$ of Fe$_{1.05}$Te in zero field and 14 T
(magnetic field along the c-axis). At high temperature, the
resistivity shows a semiconducting-like behavior, i.e.,
$\rho_{ab}$ increases slowly with decreasing temperature, whereas
$\rho_{ab}$ drops steeply below 65 K and then behaves quite
metallic. This discontinuous change in the resistivity at 65 K is
due to a structural phase transition, accompanied by magnetic
transition.\cite{Bao,Li} A thermal hysteresis of 2 K is clearly
observed in the resistivity data shown as an inset to Fig.1(a).
This is consistent with the recent neutron diffraction
measurements on Fe$_{1+y}$Te polycrystalline samples, which
indicated that the P4/nmm tetragonal structure distorts to a Pmmn
orthorhombic\cite{Bao} or a $P2_1/m$ monoclinic structure\cite{Li}
and Fe$_{1.05}$Te orders into an incommensurate magnetic state
when the temperature is lowered below $\sim$ 65 K. We have also
attempted to probe the influence of magnetic fields on the
$\rho$(T) behavior. We find that the magnetic transition
temperature is insensitive to the applied field and the
magnetoresistance is negligibly small in the AF state (less than
0.5$\%$ near 2 K at the maximum applied field of 14 T). This
behavior is different from that observed in SrFe$_2$As$_2$,
[$\rho_{ab}(8T)$-$\rho_{ab}(0T)$]/$\rho_{ab}(0T)$ reaching as high
as 20$\%$ at 10 K.\cite{Chen1} This indicates that the magnetic
coupling is much stronger in FeTe compared to other FeAs-based
systems.

In Fig. 1(b), we present the temperature dependence of magnetic
susceptibility $\chi$ in a field of 1 T with H$\parallel$ab and
H$\parallel$c-axis, respectively. Near 65 K, $\chi$ decreases
abruptly by a factor of two, indicative of the presence of
first-order phase transition. Above this temperature, $\chi$ shows
a Curie-Wiess like behavior. This is in contrast to those observed
in RFeAsO (R=La and rare earth elements) and AFe$_2$As$_2$ (A=Sr,
Ba) parent compounds above
T$_{SDW}$.\cite{Kamihara08,Chen1,GMZhang} The Curie-Wiess like
behavior could be naturally attributed to local moment formation
of the excess Fe.

\begin{figure}
\includegraphics[width=8.5cm,clip]{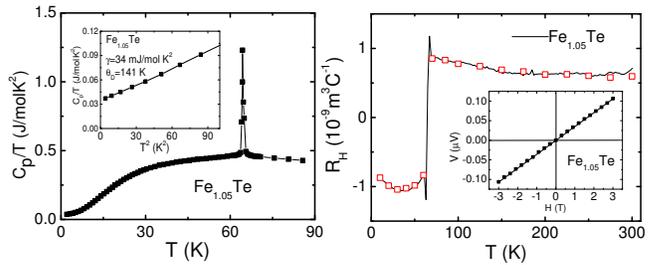}
\caption {(Color online) (a) Temperature dependence of specific
heat C for Fe$_{1.05}$Te. Inset: T$^2$ dependence of C/T in low
temperatures. (b) Temperature dependence of Hall coefficient for
Fe$_{1.05}$Te. The inset shows the transverse voltage measured at
T = 100 K is proportional to the applied magnetic field.}
\end{figure}

To get more information about the magnetic and structural phase
transition, we preformed specific heat measurement for
Fe$_{1.05}$Te. Figure 2(a) shows the temperature dependence of
specific heat C$_p$ from 2 to 90 K. We can see clearly a sharp
$\delta$-function shape peak at about 65 K with $\Delta$C $\sim$
0.8 J/molK. This is a characteristic feature of a first-order
phase transition; even though the relaxation method employed in
PPMS to measure heat capacity suppresses sharp features due to
first-order transitions. The transition temperature agrees well
with that observed in resistivity and magnetic susceptibility
measurements. It is worth noting that only one peak at around 65 K
is observed in Fe$_{1.05}$Te. This suggests the structural
transition and magnetic order occur at the same temperature. At
low temperatures, a good linear T$^2$ dependence of C$_p$/T is
observed, indicating that the specific heat C$_p$ is mainly
contributed by electrons and phonons (see inset of Fig. 2(a)). The
fit yields the electronic coefficient $\gamma$=34 mJ/mol.K$^2$,
and the Debye temperature $\theta_D$ =141 K. The electronic
coefficient is significantly larger than the values obtained from
the band structure calculations.\cite{Lu} Note that, for the
parent compound of LaFeAsO and SrFe$_2$As$_2$, the electronic
coefficients are significantly smaller than the values obtained
from the band structure calculations for non-magnetic state. It
was explained as due to a partial energy gap opening below SDW
transition which removes parts of the density of
states.\cite{Dong,Chen1} Thus, the higher value here suggests an
absence of gap opening below the transition.

Figure 2(b) shows the Hall coefficient data between 10 and 300 K
for Fe$_{1.05}$Te. The inset shows the verification of the Hall
voltage driven by magnetic field where a linear dependence of the
transverse voltage on the applied magnetic field is observed up to
3 T at 100 K. Two sets of data were presented in the main panel.
The red squares are Hall coefficient data measured by scanning
magnetic field at fixed temperature, while the solid black curve
is R$_H$ determined from two separate temperature-scan under fixed
applied magnetic field at 3 T, respectively. They show a rather
good match. In Fig. 2(b), we find that there is an abrupt sign
change in R$_H$ at 65 K, but the absolute values are almost
unchanged. Furthermore, R$_H$ in Fe$_{1.05}$Te is nearly
temperature independent even below 65 K. Those behaviors are
considerably different from the FeAs-based parent compounds. For
example, in the case of SrFe$_2$As$_2$, R$_H$ drops quickly below
T$_{SDW}$ and continuously decreases to a very large negative
value with decreasing temperature. The absolute value of R$_H$ at
2 K is about 35 times larger than that above
T$_{SDW}$.\cite{Chen1} The huge increase of the R$_H$ value is
naturally explained by the gapping of the FS which removes a large
part of free carriers. This was indeed confirmed by the optical
spectroscopy measurement.\cite{Hu} Here, R$_H$ only shows an
abrupt sign change but keeps almost the same absolute value. This
may again indicate that the FS is not gapped but caused by other
mysterious reasons, as opposed to the FeAs-based compounds.

\begin{figure}
\includegraphics[width=8.5cm,clip]{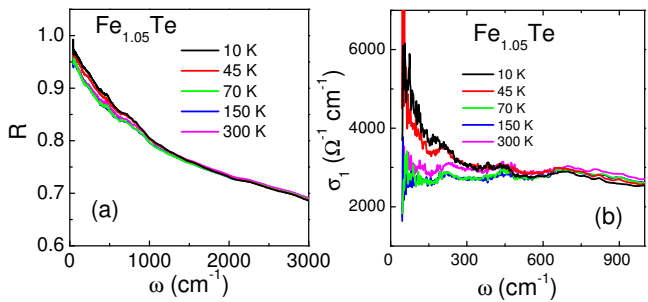}

\caption {(Color online) Optical reflectance (a) and conductivity
(b) spectra at different temperatures for Fe$_{1.05}$Te. }
\end{figure}

To further probe the carrier dynamics, we measured the optical
response of the Fe$_{1.05}$Te crystal. Figure 3 shows the
reflectance R($\omega$) and conductivity $\sigma_1(\omega$)
spectra at different temperatures. In accord with the
semiconductor-like T-dependent dc resistivity above T$_{SDW}$, we
find that the low frequency reflectance decreases slightly with
decreasing temperature from 300 to 70 K, leading to a reduction of
low-$\omega$ conductivity. It is important to note that the
conductivity spectra above T$_{SDW}$ are rather flat. Definitely,
it is not a semiconductor as $\sigma_1(\omega$) does not indicate
a presence of any semiconductor-like gap, but it is not a usual
metal as well because of the absence of a Drude-like peak. As is
well known, the width of Drude peak is linked with the scattering
rate (or inverse of the transport lifetime) of the quasiparticle,
the measurement result indicates that there is no well defined
quasiparticle with sufficiently long transport life time above
T$_{SDW}$. The charge transport is rather incoherent. Below
T$_{SDW}$, the low-$\omega$ R($\omega$) increases fast.
Correspondingly, a Drude component in $\sigma_1(\omega$) develops
quickly from the incoherent background. However, unlike the parent
compounds of LaFeAsO and AFe$_2$As$_2$ (A=Ba,Sr), there is no
partial gap formation in optical spectra below T$_{SDW}$, which is
consistent with the specific heat and Hall coefficient data
mentioned above.

All above experiment results indicate that the Fe$_{1.05}$Te
compound is very different from the LaFeAsO and AFe$_2$As$_2$
(A=Ba,Sr) parent compounds. In LaFeAsO and AFe$_2$As$_2$
(A=Ba,Sr), the charge carriers are itinerant, the SDW order is
driven by the nesting of FS, which also leads to a partial energy
gap formation and a removal of a large part of conducting
carriers. However, in Fe$_{1.05}$Te, no well-defined
quasiparticles with sufficiently long life time exist above
T$_{SDW}$, very likely due to strong magnetic scattering from the
excess Fe ions, and furthermore, no clear gap formation is seen
below T$_{SDW}$. Those behaviors might be understood from a local
moment picture with a complex frustrations,\cite{Lu} but a recent
density functional calculations based on itinerant framework
indicate that the doping effect and the large magnetic moment
induced by excess Fe could cause a size mismatch between the
electron and hole Fermi surfaces and a disappearance of
commensurate SDW order.\cite{Zhang} The cooperative effect of
excess Fe and the electrons in the FeTe layers may give such
complicated behavior although the driving force is the FS nesting
tendency which is particularly strong in FeTe.\cite{Zhang}

\begin{figure}
\includegraphics[width=8.5cm,clip]{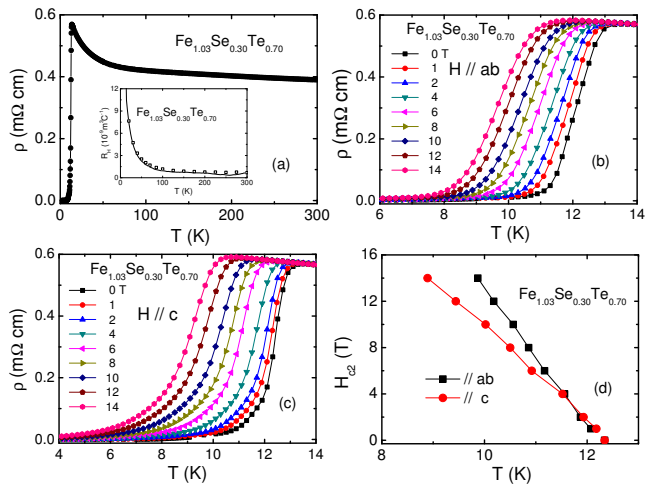}
\caption {(Color online) (a) Temperature dependence of the
in-plane electrical resistivity for
Fe$_{1.03}$Se$_{0.30}$Te$_{0.70}$ at zero field. Inset:
Temperature dependence of Hall coefficient for
Fe$_{1.03}$Se$_{0.30}$Te$_{0.70}$.  (b)and (c) Temperature
dependence of the in-plane electrical resistivity for
Fe$_{1.03}$Se$_{0.30}$Te$_{0.70}$ in low temperature region at
fixed fields up to 14T for H $\parallel$ab plane and H
$\parallel$c-axis, respectively. (d)H$_{c2}$(T) plot for H
$\parallel$ab plane (closed square) and H $\parallel$c-axis
(closed circle), respectively.}
\end{figure}

Finally we present the physical properties of a superconducting
single crystal Fe$_{1.03}$Se$_{0.30}$Te$_{0.70}$. In such Se doped
sample, the magnetism/structure instability is suppressed and
superconductivity appears instead. Figure 4(a) shows the
temperature dependence of resistivity $\rho_{ab}$ on the single
crystal Fe$_{1.03}$Se$_{0.30}$Te$_{0.70}$ with I$\parallel$ab at
zero field. $\rho_{ab}$ increases with decreasing T and shows a
sharp drop to zero at about 11 K, indicating a superconducting
transition. The Hall coefficient R$_{H}$, as shown in the inset of
Fig. 4(a), is positive, suggesting dominantly hole-type conducting
carriers. Figure 4(b) and Figure 4(c) show $\rho_{ab}$(T) for
Fe$_{1.03}$Se$_{0.30}$Te$_{0.70}$ in external magnetic fields up
to 14 T within ab plane and along c-axis, respectively. We can see
the superconducting transition is broadened slightly in magnetic
fields up to 14T. The behavior is rather different from
polycrystalline LaOFeAs where the superconducting transition is
broadened strongly in magnetic fields.\cite{Chen2} Figure 4(d)
shows H$_{c2}$-T$_c$ curves for both H$\parallel$ab and
H$\parallel$c, respectively, where T$_c$ is defined by a criterion
of 50$\%$ of normal state resistivity. The curves H$_{c2}$(T) are
very steep with slopes -dH$_{c2}^{ab}$ /dT $\mid _{T_c}$=5.96 T/K
for H$\parallel$ab and -dH$_{c2}^{c}$/dT$\mid_{T_c}$=3.69 T/K for
H$\parallel$c. Using the Werthamer-Helfand-Hohenberg
formula\cite{WHH} H$_{c2}$(0)=-0.69(dH$_{c2}$/dt)T$_c$ and taking
T$_c$=12.4 K, the upper critical fields are estimated as
H$_{c2}^{ab}$=51 T and H$_{c2}^{c}$=31 T, respectively. It is
important to note that those values are very large, comparable to
the values found for F-doped LaFeAsO with T$_c$ higher than 20
K.\cite{Chen2}

It has been well known that, for the two-gap superconductivity in
the dirty limit, the impurity scattering could strongly enhance
the upper critical field H$_{c2}$.\cite{Gurevich} Such a trend is
clearly observed for MgB$_2$, a dirty two-gap superconductor,
where the H$_{c2}$ increases remarkably with the increase of
resistivity by alloying MgB$_2$ with nonmagnetic
impurities.\cite{Gurevich} Then, the very high H$_{c2}$ observed
in Fe$_{1.03}$Se$_{0.30}$Te$_{0.70}$ could be reasonably
attributed to the strong impurity scattering effect from the
randomly distributed excess Fe in such a new multiple band
superconductor. The impurity scattering, particularly in the case
of Fe$^{+}$ with a magnetic moment, would also suppresses T$_c$.
This could be the reason why T$_c$ is not so high, although
density functional calculations indicated a higher strength of SDW
order in stoichiometric FeTe, and therefore expected a higher
T$_c$ after the suppression of SDW order.\cite{Sybedi,Zhang}

To conclude, the electronic properties of single-crystalline
Fe$_{1.05}$Te and Fe$_{1.03}$Se$_{0.30}$Te$_{0.70}$ were studied.
The excess Fe ions in those compounds dramatically modify their
properties. For Fe$_{1.05}$Te above T$_{SDW}$, we could not
identify well-defined quasiparticles with sufficiently long life
time. The charge transport is rather incoherent. Below T$_{SDW}$,
a small Drude weight develops from the incoherent background. But
unlike LaFeAsO and SrFe$_2$As$_2$, there is no clear sign of gap
opening from optical, Hall coefficient and low-T specific heat
measurement results. Strong impurity scattering effect seems to
exist also in Se-doped superconducting sample, which leads to a
relatively low T$_c$ but a rather high upper critical field in
such a new mutiple band superconductor.

This work is supported by the NSFC, CAS, and the 973 project of
the MOST of China.

\end{document}